\documentclass[nonacm,sigplan]{acmart}

\usepackage[]{hyperref}

\usepackage{multirow}
\usepackage{listings, listings-rust}
\usepackage[htt]{hyphenat}
\usepackage{orcidlink}
\newcommand{\code}[1]{\texttt{#1}}

\begin{document}

\title{A Case Study on Model Checking and Runtime Verification for Awkernel}

\author{Akira Hasegawa\orcidlink{0009-0003-6853-3989}}
\affiliation{%
  \institution{Japan Advanced Institute of Science and Technology}
  \city{Nomi}
  \country{Japan}
}
\email{akira.hasegawa@jaist.ac.jp}
\author{Ryuta Kambe}
\affiliation{%
  \institution{TIER IV, Inc.}
  \city{Tokyo}
  \country{Japan}
}
\email{ryuta.kambe@tier4.jp}
\author{Toshiaki Aoki\orcidlink{0000-0002-1209-6375}}
\affiliation{%
  \institution{Japan Advanced Institute of Science and Technology}
  \city{Nomi}
  \country{Japan}
}
\email{toshiaki@jaist.ac.jp}
\author{Yuuki Takano\orcidlink{0000-0002-6888-9024}}
\affiliation{%
  \institution{TIER IV, Inc.}
  \city{Tokyo}
  \country{Japan}
}
\email{yuuki.takano@tier4.jp}

\begin{abstract}
  In operating system development, concurrency poses significant challenges.
  It is difficult for humans to manually review concurrent behaviors or to write test cases covering all possible executions, often resulting in critical bugs.
  Preemption in schedulers serves as a typical example.
  
  This paper proposes a development method for concurrent software, such as schedulers. Our method incorporates model checking as an aid for tracing code, simplifying the analysis of concurrent behavior; we refer to this as model checking-assisted code review.
  While this approach aids in tracing behaviors, the accuracy of the results is limited because of the semantics gap between the modeling language and the programming language. Therefore, we also introduce runtime verification to address this limitation in model checking-assisted code review.
  
  We applied our approach to a real-world operating system, Awkernel, as a case study.
  This new operating system, currently under development for autonomous driving, is designed for preemptive task execution using asynchronous functions in Rust. 
  After implementing our method, we identified several bugs that are difficult to detect through manual reviews or simple tests.
  \end{abstract}
  
  \maketitle % should come after the abstract
  \pagestyle{plain} % should come right after \maketitle
  
  \section{Introduction}
  Autonomous driving technology has recently rapidly progressed and gained global attention.
  While there are various methods for achieving autonomous driving, most modern autonomous vehicles are equipped with numerous sensors sensors, including LiDAR and cameras.
  Because of the volume of sensors, these vehicles must process several gigabits of data in real time, necessitating high throughput from their operating systems.
  
  Additionally, as the procedures for processing sensor data are asynchronous, operating systems must facilitate their implementation and ensure effective execution.
  Moreover, these procedures should be preemptive to enable real-time scheduling.
  
  Given these requirements, the scheduler's behavior in such operating systems exhibits significant interleavings, a consequence of increasing throughput. Additionally, because of the asynchronous nature, processes running on the OS may not executed at the fixed intervals but can occur at virtually any moment, leading to a vast range of possible behaviors.
  
  Developing a scheduling mechanism for such operating systems is challenging, as the number of potential paths exceeds human capacity.
  Even with many people assigned to review the scheduler's code carefully, it is likely that we will miss the paths that lead to bugs.
  Even if we identify the bugs and create a patch to fix them, we cannot be confident that the patch will effectively resolve the issues or that similar bugs will never occur in the updated code.
  
  How can we mitigate this problem?  This is a challenging task, but it must be addressed to create a reliable operating system for autonomous driving.
  
  To tackle this challenge, we propose a development method that incorporates model checking and runtime verification into the operating system development process.
  Our method uses model checking to make it easier to trace potential paths based on our understanding of the code.
  We first write a behavior in a modeling language that aligns with our understanding and then check the behavior that can happen if the description is correct.
  Writing a description of the behavior from scratch is difficult, as grasping the whole behavior of the code is challenging due to its concurrency and asynchronous nature.
  Therefore, our approach begins by translating the code of the operating system's implementation into a modeling language used in a model checker, line by line.
  
  Of course, there is a semantic gap between the programming and modeling languages. Thus, the translation may be flawed, and some paths can be missed during checking the model. 
  Still, it simplifies the process of tracing and contemplating potential paths in the code compared to relying solely on mental processing without model checking. 
  We will refer to the model created by this translation as \textit{a code review model}.
  
  To address the limitations of the code review model caused by the semantics gap, we also introduced runtime verification to the development process. Although runtime verification does not guarantee exhaustive checking, it is useful for efficiently analyzing the actual behavior of the implementation.

  In Section \ref{sec:related-work}, we will discuss related work. In Section \ref{sec:awkernel}, we will introduce the target system of our proposed method. In Section \ref{sec:mcacr}, we will explain how we utilized model checking for code reviews in the development of Awkernel. In Section \ref{sec:rv}, we will describe how we used runtime verification to address the drawbacks of model checking. In Section \ref{sec:result}, we will present the results of applying our approach to Awkernel and identifying bugs. In Section \ref{sec:lessons-learned}, we will share the lessons learned. Finally, we will conclude the paper in Section \ref{sec:conclusion}.

\section{Related Work}
\label{sec:related-work}
Formal methods have been widely used to verify operating systems.
One approach involves theorem proving~\cite{sel4}; seL4 is an operating system developed while verifying the correctness of theorem proving.
Although their approach seems solid, it is time-consuming. They indicated that they would require six person-years to redo a similar verification for a new kernel using the same methodology. Since we did not have enough team members to apply their approach and time was limited, it was impractical for our situation.
Moreover, we wanted to check task state transitions in terms of their concurrent nature. 
However, verifying such properties with theorem proving is more challenging.
Our method is less confident than seL4's approach but it is much lighter and can be completed in around three person-months.

Another work that uses formal methods to verify the behavior of an operating system exists.
de Oliveira et al.~\cite{Oliveira2019} proposed a method for conducting runtime verification to check thread behavior 
in the PREEMPT\_RT Linux kernel.
First, they created a model of the behavior of thread behavior in the PREEMPT\_RT Linux kernel using automata while checking logs of actual execution with \code{perf} and obtained reviews from Linux kernel maintainers~\cite{OliveiraModeling1,OliveiraModeling2,OliveiraModeling3}.
Second, they converted the model into a C data structure and embedded it into the Linux kernel as a module~\cite{Oliveira2019}. They then verified that the thread behavior in the Linux kernel aligns with their created model.
Their method is much lighter than the seL4 approach~\cite{sel4}. 
However, runtime verification does not guarantee exhaustiveness.
In developing a scheduler, we want to identify bugs that appear only during rare execution paths; thus, runtime verification alone is unsuitable for our needs.
Our method also employs runtime verification, but we used it not as a primary verification method but as a complement to model checking. Model checking provides exhaustiveness under several assumptions, while runtime verification assesses actual behavior and the accuracy of the model checking results.

There is also another work for verification of an operating system that combines model checking and model-based testing~\cite {Aoki2017,Aoki2023}.
The researchers used the Spin model checker to verify the scheduler's functional correctness on an automotive OS compliant with the OSEK/VDX standard, called REL OS.
Based on the OSEK/VDX standard, they first created a model of the scheduling behavior in Promela, the modeling language used by the Spin model checker.
They then verified that desirable properties were satisfied in the model through model checking.
Although the properties are satisfied in the model, this does not guarantee that the actual behavior of REL OS aligns with the model's behavior, or that it satisfies the properties. Therefore, to check the conformance between the model and the actual behavior of REL OS, they generated test cases based on the model used for model checking and conducted model-based testing. They created a search tree using a log from the Spin model checker and converted it into test cases. After that, they transformed the test case to runnable test programs and executed them.
Their approach is practical even with limited time and resources. However, creating the model requires a specification that the implementation follows, such as the OSEK/VDX standard. 
Because they were not based on such specifications and were implemented while conducting testing to check whether a program works, their method cannot applied to our development approach and target.
Instead of creating a model from a standard, we manually translated code from the implementation into Promela while accepting the semantic gap between the programming language and Promela, the modeling language.
This approach allows the methods to be applied to a broader range of targets.

\begin{figure}[tbp]
  \centering
  \includegraphics[width=\linewidth]{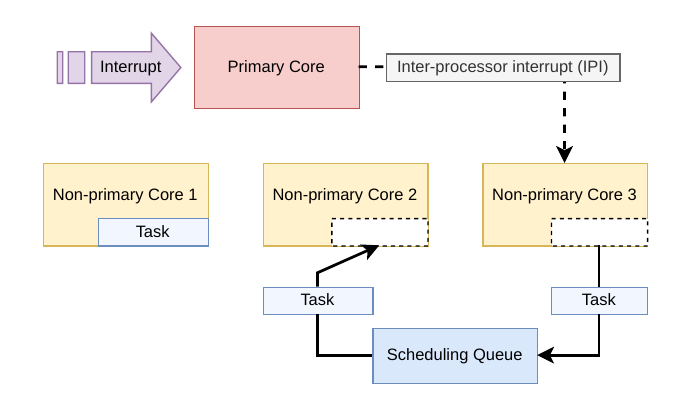}
  \caption{Overview of task scheduling and interrupt handling. The primary core does not execute tasks but handles interrupts. When an interrupt occurs, the primary core sends an inter-processor interrupt to the non-primary cores. The non-primary cores then fetch tasks from the scheduling queue and execute them.\label{fig:rust-overview}}
\end{figure}
\section{Awkernel}
\label{sec:Awkernel}

\subsection{Design Principles}
In this section, we describe the design principles of Awkernel, a new operating system for autonomous vehicles developed from scratch using Rust.
Since autonomous vehicles must process large amounts of data in real time, the OS must have high throughput and performance.
On the other hand, safety is crucial since the OS is used in autonomous vehicles, which are considered safety-critical systems.

Awkernel has been developed to achieve both high performance and safety.
It is designed to operate in a single address space, similar to unikernels, for simplicity and performance improvement.
Tasks share the same memory space.
From a safety perspective, the memory spaces used by each task must be isolated.
We achieve this by utilizing Rust's type system, similar to other Rust-based OSes~\cite{Theseus,RedLeaf}.
Thanks to this design, the OS can maintain good throughput.
Moreover, to enhance safety, it has been developed using Rust; with Rust's type system, memory spaces for each task are effectively isolated while adopting a single address space.

Although Rust's type system guarantees memory safety, it does not ensure other properties that can affect overall safety.
To improve confidence regarding safety, we introduced formal methods in the development of Awkernel. We have already checked the functional correctness of important scheduling functions using Kani~\cite{kani} and TLA+~\cite{tla}.

Preemption-related procedures have not yet been verified due to the state explosion problem. 
However, formal methods must be applied to check those processes since the number of interleavings is vast and challenging for humans alone to trace alone.

In the next section, we will describe the scheduling mechanisms that lead to the significant number of interleavings causing the state explosion problem and then explain how we used model checking to verify them.

\subsection{Asynchronous Task Execution}
\label{ssec:async-task-execution}
Since autonomous vehicles have numerous sensors, asynchronous processing is frequently employed.
If programmers are forced to use a traditional callback-based implementation, development will become more complex and error-prone~\cite{asyncawait}.
Therefore, Awkernel natively supports async/await.
Each task in Awkernel is executed as an async/await function in Rust, allowing us to employ the async/await feature for tasks, such as processing sensor data.

This mechanism enables the OS to switch tasks more efficiently than the traditional context switch; when switching tasks, the OS does not need to execute a preemption procedure that includes saving and loading contexts.
This task-switching mechanism is inspired by Embassy~\cite{embassy,EmbassyBlog}, where the OS, also developed using Rust, can switch tasks utilizing the async/await feature similar to Awkernel.
Occasionally, a task may need to wait for certain events and enter the \code{Waiting} state. 
In this case, the task returns \code{Poll::Pending} to the OS, which then switches to another task using Rust's async/await feature.
When an event arrives and the task is ready to run again, the OS resumes the task using the async/await feature.
However, the traditional context switch mechanism is still needed for real-time scheduling.
Thus, we implemented the traditional context switch mechanism for the async/await functions, making them preemptive. 
When a task wants to change its state, such as waiting for an event, it uses the async/await feature to switch tasks. Conversely, when a scheduler needs to switch a task, for instance, when a task exceeds its time slice, it utilizes the traditional context switch mechanism to switch the task.

To realize these two task-switching mechanisms, we divided the roles of CPU cores into primary and non-primary cores, as shown in Figure \ref{fig:rust-overview}.
The first CPU core, typically CPU 0, is designated as the primary core, handling timed events and scheduler operations. 
Other CPU cores are termed non-primary cores, which fetch tasks from the shared scheduling queue and execute them.

For example, when an interruption occurs externally, the primary core manages the interruption. If the primary core decides to interrupt the procedure running on some a non-primary core, it sends inter-processor interrupts (IPIs) to the specific non-primary cores related to the interruption, as shown in Figure \ref{fig:rust-overview}, and interrupts the tasks.

Similarly, when a timed event arrives, the primary core will manage it as well. A typical example is the round-robin scheduler, whose operations run on the primary core, checking whether each task's execution time exceed its time slice.
If the primary core identifies a task that exceeds its time slice, it sends an IPI to the non-primary core running the task.
The non-primary core that receives the IPI will suspend the task's execution and initiate preemption. The preempted task will be enqueued into the scheduling queue, as shown in Figure \ref{fig:rust-overview}, and may be resumed on another non-primary core.

\paragraph{Running Example} 
\label{para:running-example}

In Awkernel, each CPU core operates almost independently. The primary core continuously processes timed events and performs scheduler operations, while non-primary cores fetch tasks from the scheduling queue and execute them, as shown in Figure \ref{fig:rust-overview}.

However, the queue used for scheduling and the data structures for tasks are shared among all CPU cores. Each core accesses this shared data at different times, resulting in numerous interleavings. Consequently, the order of processing can sometimes differ from expectations, potentially leading to bugs.

Here, we present a concurrency-related bug found during Awkernel's development. The OS has a function called \code{wait()} to enqueue a task into the scheduling queue, which takes a task as an argument. Additionally, there is a delta list for handling timed events. When an item is added to the delta list, the primary core checks the events, and when one arrives, the callback function defined for the item in the delta list is called.

Assume there is a running task that adds itself to the delta list to wait for a timed event. Also, assume the callback function is \code{wake()}. In this case, we can consider the following two behaviors:

\paragraph{Behavior 1}
\begin{enumerate}
\item The task is added to the delta list.
\item The task changes its state from \code{Running} to \code{Waiting} and suspends execution.
\item The timed event arrives, and \code{wake()} is called, which enqueues the task into the scheduling queue.
\item After a while, the task is taken from the scheduling queue and resumed.
\end{enumerate}
\paragraph{Behavior 2}
\begin{enumerate}
\item The task is added to the delta list.
\item The timed event arrives, and \code{wake()} is called; however, since the task is still \code{Running}, the function call is ignored, and the task is not enqueued into the scheduling queue.
\item The task changes its state from \code{Running} to \code{Waiting} and suspends execution.
\item The task will never resume since the scheduling queue does not contain the task.
\end{enumerate}

Bugs caused by concurrency, such as the one illustrated in Behavior 2, are difficult to find with manual code reviews alone. Therefore, we used formal methods such as runtime verification and model checking for code reviews.

\subsection{Real-time Properties}
Autonomous driving vehicles require real-time properties. To ensure this, the OS must support real-time scheduling, and the tasks must be preemptive.

To make the async/await functions preemptive, we implemented the traditional context switch mechanism in Awkernel and made the async/await functions preemptive.
When a preemption occurs, the OS interrupts the running task and switches to the procedure for handling the preemption, such as saving the task's context and loading or creating the context of the next task.

\section{Model Checking}
\label{sec:mcacr}
\subsection{Approach}
In this section, we will show how we used model checking for code reviews in the development of Awkernel.
During the development, we used the Spin model checker~\cite{spin} for reviews.
The Spin model checker is a widely used tool for verification concurrent systems, and in our experience, its execution is faster than the other model checkers.

Although the Spin model checker requires a description of the target behavior in Promela, a modeling language used by Spin, the concurrent nature of the scheduling mechanism made the behavior uncertain for us and difficult to describe.
Thus, we manually translated the Rust code into Promela line by line while minimizing behavioral changes caused by semantics differences.
We will refer to the translated Promela description as \textit{the code review model}, with details are provided in Section \ref{ssec:line-by-line}.

Although Awkernel supports various types of schedulers that can be added flexibly,
when writing this paper, Awkernel includes two schedulers: a round-robin scheduler and a first-in, first-out (FIFO) scheduler.
Since the round-robin scheduler operates in the same way as the FIFO scheduler when preemption does not occur, 
we omitted the FIFO scheduler from the code review model and focused on the round-robin scheduler.

When manually translating Rust code to Promela, hardware-related code is particularly challenging.
For example, when a non-primary core receives an IPI, it immediately suspends the execution of the regular operation procedure and begins the preemption procedure.
This behavior cannot be modeled directly by translating the code.
Moreover, the preemption procedure itself is also challenging to model in Promela; each task's context must be saved, and when the task resumes, the context must be loaded and restored.
Section \ref{ssec:preemption} describes our approach to addressing these issues.

Another issue arises regarding the handling of time within the code review model.
In Awkernel, each core shares and modifies the data used for scheduling at various times, creating many interleavings. Hence, if we introduce a time integer variable into the code review model, the state explosion problem will occur.
For example, a task running for a long time will be preempted in round-robin scheduling. If we introduce the time concept directly to the code review model to maintain and check the execution time of each task, the sequence of all variables is considered a state in the Spin model checker, which leads to an increase in search depth and results in state explosion.
To avoid this, we used an over-approximation of the time concept to simplify the code review model.
We will explain how we modeled the time-related procedure in the Promela model in Section \ref{ssec:time}.

Translating the Rust code to Promela is insufficient to check the scheduling mechanism's behavior.
We must prepare tasks running on the schedulers to observe practical operations in model checking. 
Since describing all possible behaviors of tasks is nearly impossible, we have also approximated task behavior.
Our approach to approximating tasks behaviors is described in Section \ref{ssec:task}.

In the development of Awkernel, model checking techniques have been valuable for identifying several bugs.
The bugs we found and how we fixed them will be described in Section \ref{ssec:mc-found-bugs}.

\subsection{Line-by-Line Translation}
\label{ssec:line-by-line}
To create the code review model, we first translated the Rust code of Awkernel into Promela line by line. For example, the code snippet shown in Figure \ref{fig:non-primary-example} was translated into Promela as shown in Figure \ref{fig:promela-non-primary-example}. 

To create the model of Awkernel correctly, we need to consider the differences in semantics between Rust and Promela, including hardware-related behaviors.
However, this task is extremely time-consuming; we must clarify the details of Rust's semantics and how the hardware interprets and executes them concurrently.
To leverage model checking within a limited development time, we balanced our approach by translating the implementation into Promela based on our understanding. In other words, the code review model does not represent the actual behavior of the implementation; it is constructed based on the developers design choices.

\begin{figure}[tbp]
\begin{lstlisting}
proctype run_main(byte __ctx_id) provided (CONTEXT_STATE[__ctx_id] == CtxActive) {
    ......
    RAWCPUIDS[awkernel_lib_cpu_id()] = awkernel_lib_raw_cpu_id();

    main_loop:
    do
    :: get_next_task(option_task_id) ->
        assert(interrupt_flag[awkernel_lib_cpu_id()] == false);
        if
        :: OPT_IS_SOME(option_task_id) ->
            _tid = option_task_id.value;
            take_preempt_context(_tid, option_context);
            if
            :: OPT_IS_SOME(option_context) ->
                yield_and_pool(option_context.value);
                goto main_loop
            :: OPT_IS_NONE(option_context)
            fi;
            ......
\end{lstlisting}
\caption{Code snippet of procedure that runs on non-primary cores in Awkernel.\label{fig:non-primary-example}}
\end{figure}

\begin{figure}[tbp]
\begin{lstlisting}
 pub fn run_main() {
   CPUID_TO_RAWCPUID[awkernel_lib::cpu::cpu_id()]
       .store(awkernel_lib::cpu::raw_cpu_id(), Ordering::Relaxed);
   ......

   loop {
       if let Some(task) = get_next_task() {
           #[cfg(not(feature = "no_preempt"))]
           {
               let mut node = MCSNode::new();
               let mut info = task.info.lock(&mut node);

               if let Some(ctx) = info.take_preempt_context() {
                   info.update_last_executed();
                   drop(info);

                   unsafe { preempt::yield_and_pool(ctx) };
                   continue;
               }
           }
           ......
\end{lstlisting}
\caption{Code snippet of the Promela description that corresponds to the procedure running on non-primary cores in Awkernel.\label{fig:promela-non-primary-example}}
\end{figure}

Let us look at how we converted Rust to Promela through several examples.
\paragraph{Vec/VecDeque} In Promela, there is no syntax for dynamic arrays like \code{Vec} and \code{VecDeque} in Rust.
In the implementation of Awkernel, \code{Vec}/\code{VecDeque} are used as types of queues for scheduling.
Although modeling the behavior of these types directly in Promela is difficult, we know that the maximum length of the queues is the number of tasks. Therefore, we replaced \code{Vec} and \code{VecDeque} with a static array, where the length matches the number of tasks in the code review model.
Futhermore, \code{VecDeque} has several methods, such as \code{push\_back}, \code{pop\_front}.
However, we are unsure about the details of how these methods are executed at the hardware level. Hence, we assumed them to be atomic operations in Promela; for instance, we assumed that dequeuing a task from a queue, removing the task, and assigning a return value to the register are atomically, with no preemption occurring between these actions

\paragraph{Option/Result} Promela does not have types that correspond to the \code{Option} and \code{Result} types in Rust. Therefore, we treated them as structures consisting of two members: a label and a value. For example, we represent \code{x = Some(3)} as \code{x.label = Some; x.value = 3} in Promela. Additionally, since we are unsure how their methods are executed on hardware, we cannot determine if, for example, \code{x. label = Some; x.value = 3} are executed atomically or if interruptions can happen between these two statements. Thus, similar to the methods of \code{VecDequeue}, we assumed that statements related to \code{Option} and \code{Result} types are executed atomically. 

As shown above, we ignore the details of the implementation's behavior and represent them with atomic operations. Therefore, if the timing of operations related to such types is critical, we may overlook timing-related bugs. However, we can still identify timing bugs related to function calls, as illustrated in Section \ref{ssec:mc-found-bugs}.

\subsection{Modeling of Preemption}
\label{ssec:preemption}
As discussed in the previous section, most of the code review model is created through translation. However, these translations alone are insufficient for the model to function as the actual implementation does; the translated code does not include a preemption mechanism, which must also be modeled to replicate the behavior of Awkernel.

In Awkernel, when preemption occurs, the non-primary core suspends the execution of the running task and saves the thread context at that moment. The non-primary core then performs a context switch. When the task is resumed, the non-primary core loads the saved context and resumes the task from where it was previously suspended.
There is a significant issue with describing this behavior in Promela: Promela does not have the ability to save and load a process's state.

\begin{figure}[tbp]
  \centering
  \includegraphics[width=\linewidth]{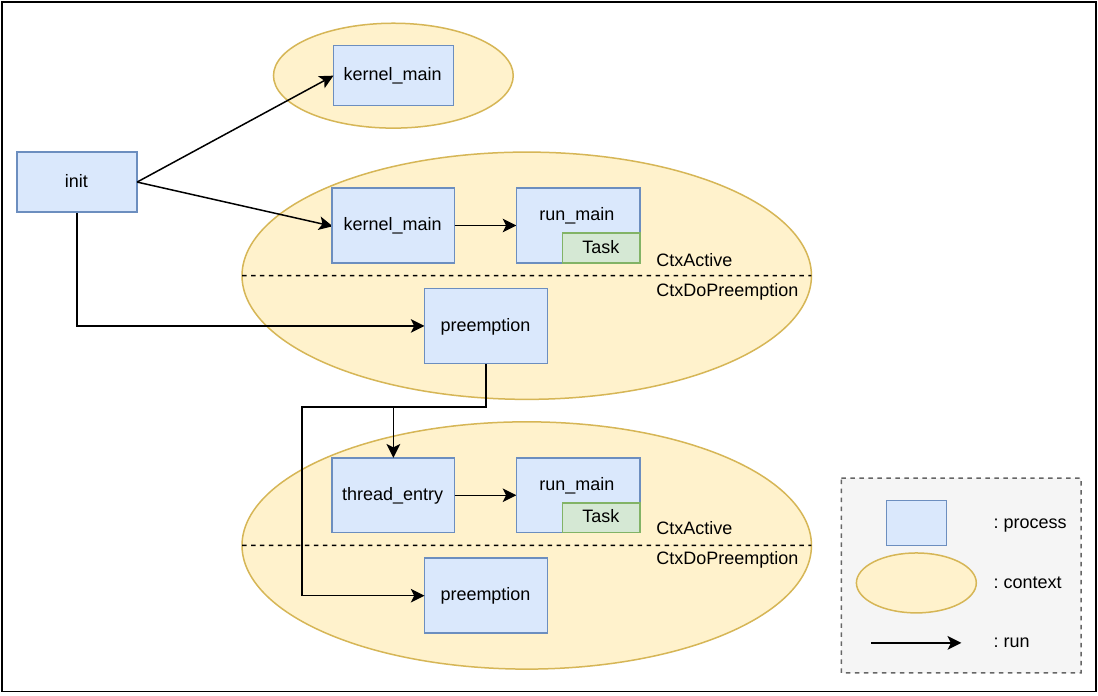}
  \caption{Overview of code review model design that is used in the model checking.\label{fig:spin-overview}}
\end{figure}

To overcome this issue, we divided one task in Awkernel into multiple tasks, as shown in Figure \ref{fig:spin-overview}, and introduced \code{provided} statements.
\code{provided} is a primitive of Promela that allows enable/disable a process based on a condition as follows:
\begin{lstlisting} 
proctype run_main(byte __ctx_id) provided (CONTEXT_STATE[__ctx_id] == CtxActive) { ...  }
\end{lstlisting} 
In this example, when the context state becomes \code{CtxActive}, the condition is satisfied, and the process \code{run\_main} will execute. 
Conversely, when the constraint is not satisfied, the process is disabled; that is, the model checker will not be selected or executed, and the state will be retained until the constraint is satisfied again. 

To specify an active process, we defined the context states as follows: \code{CtxNotInitialized}, \code{CtxActive}, \code{CtxDoPreemption}, \code{CtxPreempted}, and \code{CtxInThreadPool}. Initially, the context state is set to \code{CtxNotInitialized}. When a task begins execution, the state changes to \code{CtxActive}. 
When preemption occurs, the state changes to \code{CtxDoPreemption}, and a process corresponding to the preemption operations is enabled; simultaneously, the process corresponding to the regular operation is disabled, which corresponds to saving the task's progress. When a preempted task is resumed, the task's state reverts to \code{CtxActive}, and the process corresponding to the regular operation is enabled, allowing the resumed task to begin operations from the point just before preemption. We will not elaborate on the details of each context state here.

By introducing this mechanism, we have successfully emulated the preemption sequence of Awkernel in the code review model. An overview of the code review model after introducing this mechanism is shown in Figure~\ref{fig:spin-overview}. In the figure, each yellow circle represents an individual task; the top circle illustrates the procedure for the primary core, while the others represent tasks running on non-primary cores. The blue squares represent the Promela processes. During the regular operation, the top half of each circle is executed. When preemption occurs, the Promela processes in the top half are disabled, and the bottom half is activated and executed.

\subsection{Modeling of Time}
\label{ssec:time}
Modeling the behavior of a round-robin scheduler and the tasks requires introducing the concept of time. When a running task has exceeded its time slice, the round-robin scheduler must send an IPI to the non-primary core that is running the task. However, Spin does not incorporate the concept of time and features to handle it.

The most naive way to handle time is to introduce a variable to represent it. If we cycle the variable's value within a range, for example, counting from 0 to 9 and then back to 0, we can count the continuous running time of each task. However, this approach may lead to the state explosion problem. 

Therefore, we decided not to handle the time explicitly. Instead, to model time-related operations without the variable, we over-approximated those operations: the round-robin scheduler can send the IPI for preemption at any time. By doing so, the model checker will search for the paths that occur in the execution sequence by the implementation without introducing time variables.

However, this approximation allows for sequences where the scheduler continuously sends IPIs to all non-primary cores and where preemption occurs alternately, which would not happen in actual implementation. To exclude this sequence from the search space of model checking, we must introduce a flag to prevent continuous and infinite sending of IPIs. Once preemption occurs, the flag ensures that the next preemption cannot happen until one of the tasks completes its process.

\subsection{Modeling of Tasks}
\label{ssec:task}
To verify the scheduler's practical and meaningful behavior, it is necessary to provide tasks for execution. However, since the types of tasks and their behaviors cannot be fully known during the verification phase, we approximate a typical task's behavior. Over-approximating task behavior to cover all possible actions can easily lead to state explosion, so we chose to under-approximate the behavior by categorizing tasks into two types. These types are pre-configured for each task during model checking.

\begin{itemize}
  \item Light Task: a task that terminates quickly and cannot be preempted by the round-robin scheduler.
  \item Heavy Task: a task that takes longer to terminate, is preempted multiple times by the round-robin scheduler, and returns \code{Poll::Pending} to wait for certain events.
\end{itemize}

By defining these two types of tasks, we can identify bugs related to the timing of preemption and the 
return of \code{Poll::Pending}.
Additionally, we can detect issues that arises only when a task completes quickly. Furthermore, by selecting Light Tasks, we can reduce the search space and the time required for model checking.
Thus, if a single Heavy Task is sufficient to reproduce a bug, we can set the remaining tasks as Light Tasks to
reproduce it more quickly and efficiently.

In our Promela description, the number of generated tasks, each task type, and the order of the generation are fixed in advance by pre-configured parameters. Specifying the number of generated tasks seems sufficient; the others can be set using non-deterministic choices in model checking. However, doing so results in a state space that is too large, leading to out-of-memory errors on the computer executing model checking.
To mitigate the size of the state space, we separate it by parameters and conduct model checking for each set of parameters individually.

When conducting model checking of the code review model, we pre-configure the number of tasks to be generated, each task's type, and the order of their generation. To check all patterns exhaustively, we can set these parameters externally and then run multiple model checking processes.
Attempting to check all patterns in a single execution would lead to out-of-memory errors.
By separating the parameter sets and running model checking individually for each, we can mitigate the impact of state explosion.

\section{Runtime Verification}
\label{sec:rv}
\subsection{Approach}
Although we found that model checking is helpful for reviews, it may differ from actual behavior due to the semantic gap between Rust and Promela and misunderstandings of Rust program behavior. To bridge this gap and rectify such misunderstandings, we employed runtime verification. Runtime verification is a formal method that checks whether the runtime behavior of the target program satisfies a formal specification. In this process, the formal specification is converted into a deterministic finite automaton (DFA) and embedded into the target program's code, creating a monitor. When the monitor detects a specification violation, it raises an alert.

Runtime verification is beneficial for testing the behavior of concurrent programs. When the target program has many interleavings, the execution sequence varies with each run, making a single test run insufficient. It is impossible to know in advance how many times we need to execute tests. Since runtime verification can operate indefinitely, even though exhaustiveness of execution for possible paths is not guaranteed, it allows us to test the behavior of the target program multiple times more efficiently than manual testing.

In our work, we created a DFA representing task transitions, converted it to a monitor, and embedded it into the code of Awkernel.
Consequently, we successfully identified a concurrency bug that would be difficult for humans to find, even with model checking.

\subsection{Modeling Task Transitions}
\begin{figure}[tbp]
  \centering
  \includegraphics[width=\linewidth]{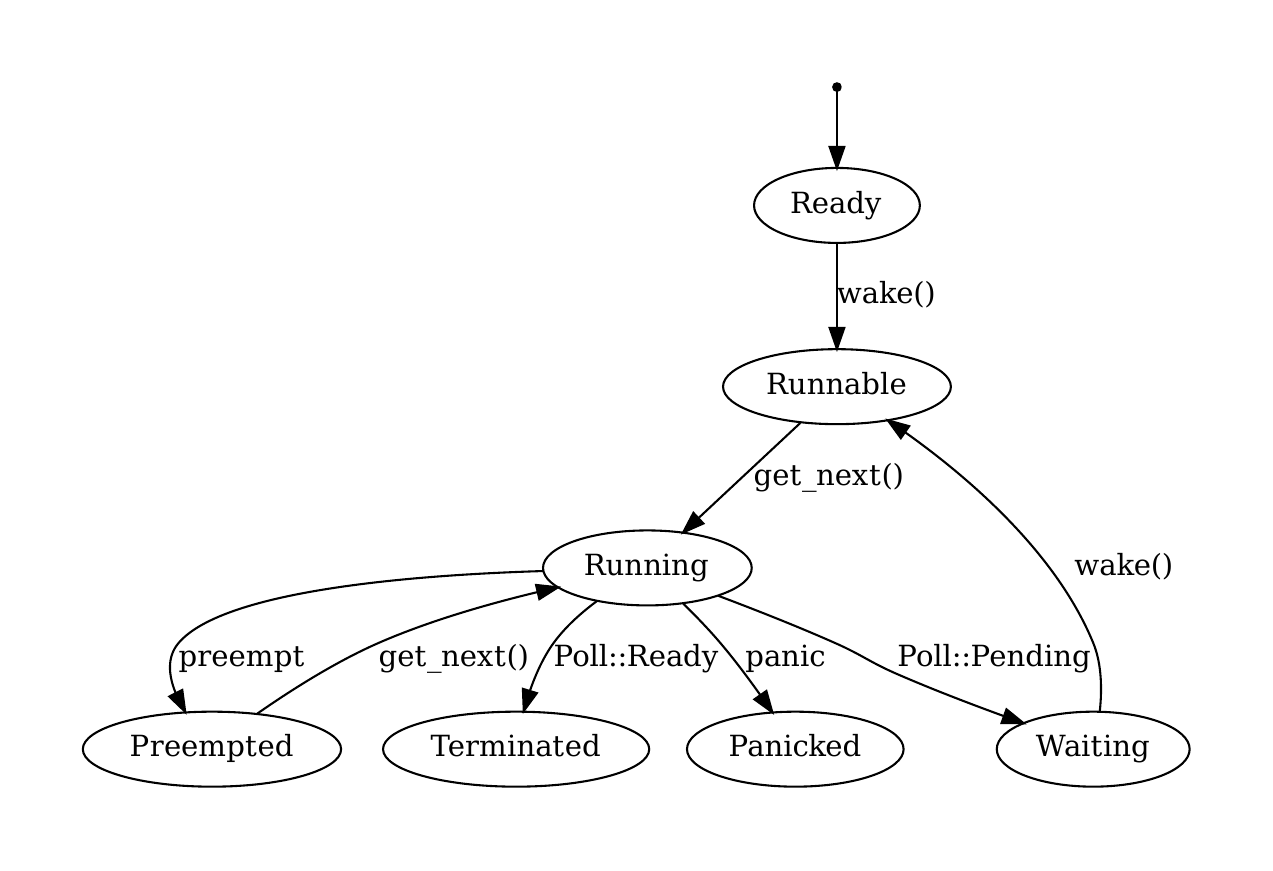}
  \caption{First diagram about task transitions in Awkernel; the diagram was created as the first model representing our understanding of the task transition. It was used as the base of the monitor while conducting runtime verification.\label{fig:task-state-transition}}
\end{figure}

Although we introduced formal methods in the development of Awkernel, we did not have a concrete model representing the actual task state transitions. 
Although a rough model, shown in Figure \ref{fig:task-state-transition}, existed, it had not been tested for conformance between the model and the actual behavior.
Hence, we began using this model as the basis for a monitor of runtime verification.

However, the actual behavior was not as straightforward as the model shown in Figure \ref{fig:task-state-transition} due to a flag called \code{need\_sched}.
In Awkernel, each task has a flag \code{need\_sched} that allows \code{Waiting} tasks to operate without timing bugs. Specifically, when the timer event arrives before the task becomes \code{Waiting}, the flag is set, and when the task transitions from \code{Running} to \code{Waiting}, the flag is checked. If the flag is set, the task will be enqueued into the scheduling queue immediately.

We believed that the flag should be tracked in runtime verification, so we added it to the model.
The process of refining the model was conducted with input from the developer of the task scheduling mechanism to ensure the model's behavior aligns with our intentions. However, we found the updated model still lacked transitions through runtime verification.

\subsection{Embedding the Monitor into Awkernel}
\begin{figure}[tbp]
  \centering
  \includegraphics[width=\linewidth]{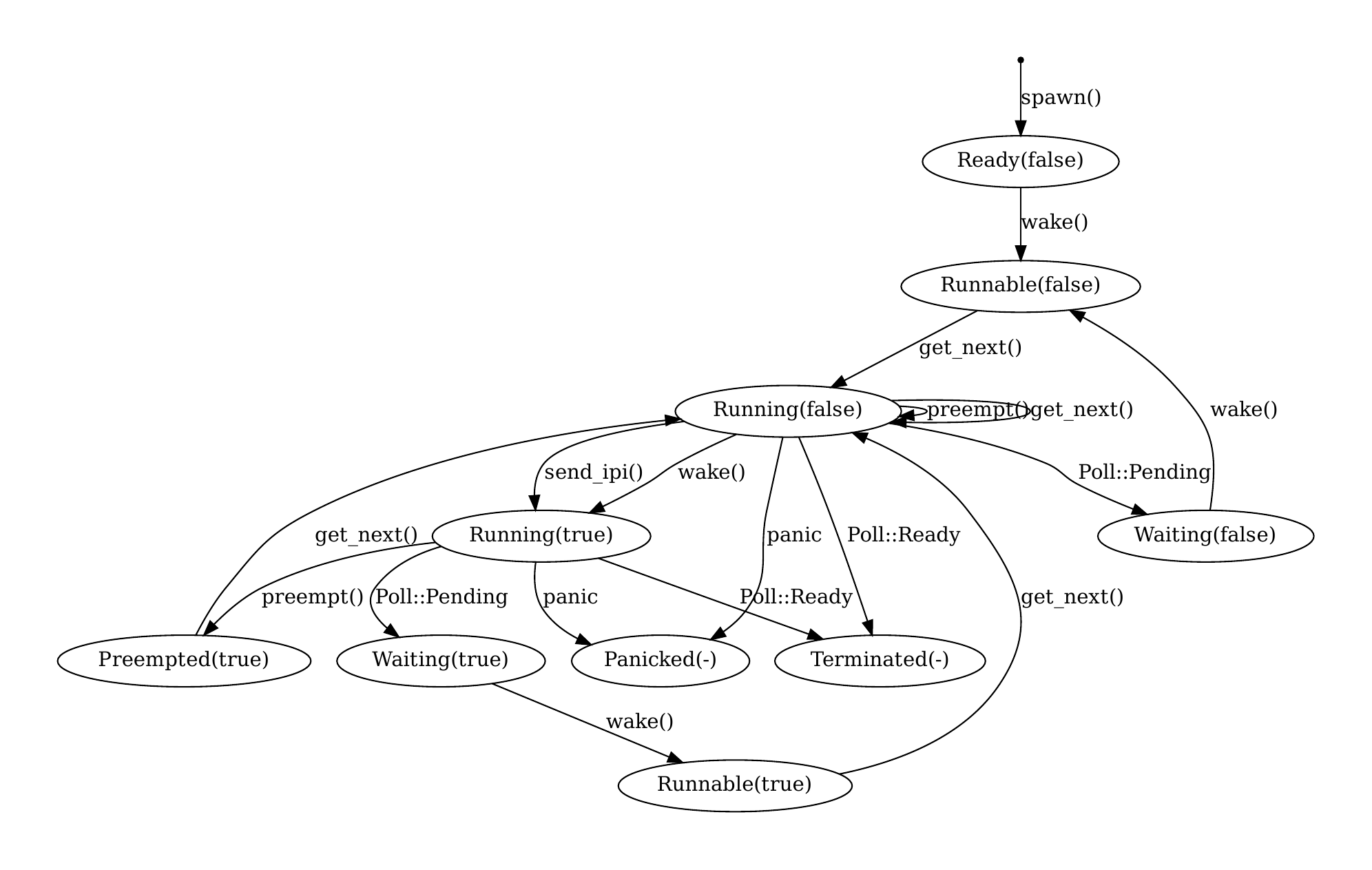}
  \caption{The final model representing task transitions in Awkernel, obtained during the continuous process of runtime verification and appending the discovered transitions\label{fig:rv-monitor}}
\end{figure}

To verify that actual task transitions follow the model, we first need to define events and states to be traced and then implement the corresponding monitoring code in Rust.

Figure \ref{fig:rv-monitor} shows the defined events and states. In the model, events are defined as function calls that change a task state. For example, when a task is enqueued in the scheduling queue, \code{wake()} function is called, and this event corresponds to the transition \code{Poll::Ready} in Figure \ref{fig:rv-monitor}.
Another example is when the task finishes execution and becomes \code{Terminated}; it will return \code{Poll::Ready}, which corresponds to the transition \code{Poll::Ready} in Figure \ref{fig:rv-monitor}.

We defined the states using a pair of variables, \code{state} and \code{need\_sched}, in the model. For example, when the \code{wake()} function is called for a running task, \code{need\_sched} will be set, and in this case,
the state will change from \code{Running(False)} to \code{Running(True)} with the transition \code{wake()}, as shown in Figure \ref{fig:rv-monitor}.
In the model, we ignored the \code{need\_sched} flag for \code{Terminated} and \code{Panicked} since they will not transition anymore. This means the flag will not be used, so we can omit it and represent the flag value as a symbol indicating ``don't care'' in Figure \ref{fig:rv-monitor}.

After defining the states and transitions, we implemented the monitor, as shown in Figure \ref{code:rv-monitor}. The monitor contains the model from \ref{fig:rv-monitor} for each task and maintains the current state of each task in the variable \code{MODELS}, as depicted in Figure \ref{code:rv-monitor}.

When a function call corresponding to one of the events occurs, we change the state of the target task in the variable \code{MODELS} accordingly. For example, when \code{spawn()} is called, a new task will be created, with the task ID assigned to the variable \code{id}, which is set to the return value of \code{tasks.spawn()}, and its state will be initialized. In this case, for the monitor, we will initialize a new model for the the task and add it to the variable \code{MODELS} as shown in L.14-L24 of Figure \ref{code:rv-hook}. 

\begin{figure}[tbp]
  \centering
  \begin{lstlisting}
  #[derive(Clone)]
  pub struct Model<State, Transition>
  where
     State: Display + Clone + Eq + Debug,
     Transition: Display + Clone + Eq + Debug,
  {
     states: Vec<State>,
     transitions: Vec<(State, Transition, State)>,
     current_state: State,
  }

  pub type TaskModel = Model<TaskModelState, Event>;

  pub static MODELS: Mutex<BTreeMap<u32, crate::task::TaskModel>> = Mutex::new(BTreeMap::new());
  \end{lstlisting}
  \caption{The code snippet of the monitor for runtime verification.\label{code:rv-monitor}}
\end{figure}

\begin{figure}[tbp]
  \centering
  \begin{lstlisting}
  pub fn spawn(
      name: Cow<'static, str>,
      future: impl Future<Output = TaskResult> + 'static + Send,
      sched_type: SchedulerType,
   ) -> u32 {
    ......
      let scheduler = get_scheduler(sched_type);
      let id = tasks.spawn(name, future.fuse(), scheduler, sched_type);
      #[cfg(feature = "runtime_verification")]
      {
        ......
          let mut models = runtime_verification::MODELS.lock(&mut node);
          let mut model = runtime_verification::task::new_task_model();
          model.next(&runtime_verification::event::Event::Spawn).unwrap();
          models.insert(id, model);
      }
      tasks.wake(id);
      id
   }
  \end{lstlisting}
\caption{The code snippet of the hook to the function for task spawning for runtime verification.\label{code:rv-hook}}
\end{figure}

\section{Results}
\label{sec:result}
\subsection{Code Review Model}
\label{ssec:loc}
The code review model was created by translating the implementation of a kernel into Promela, the modeling language used by the Spin model checker. Unlike Rust, the modern import feature does not exist in Promela, but it supports a directive similar to that in C. Consequently, some files had to be split into two parts, such as \code{task} and \code{preempt}.

The lines of code for each file and the reference file of Rust are shown in Table~\ref{tbl:loc}. Generally, each file name corresponds directly to an actual Rust file; for example, \code{task\_part1} and \code{task\_part2} correspond to \code{task.rs} in Awkernel. The exceptions are \code{verification}, \code{rust\_primitives}, and \code{main}. The \code{verification} file defines parameters for model checking, such as the number and types of tasks. The \code{rust\_primitives} file includes macros to bridge the semantic gap between Rust and Promela. Lastly, \code{main} defines the initialization sequence necessary to execute the model.

 \begin{table}
 \begin{tabular}{|ll|l|}\hline
 \multicolumn{2}{|l|}{File Name}                        & LOC \\ \hline\hline
 \multicolumn{2}{|l|}{main}                             & 83  \\ \hline
 \multicolumn{2}{|l|}{kernel\_main}                     & 32  \\ \hline
 \multicolumn{1}{|l|}{\multirow{2}{*}{task}}    & part1 & 73  \\ \cline{2-3} 
 \multicolumn{1}{|l|}{}                         & part2 & 231 \\ \hline
 \multicolumn{1}{|l|}{\multirow{2}{*}{preempt}} & part1 & 76  \\ \cline{2-3} 
 \multicolumn{1}{|l|}{}                         & part2 & 175 \\ \hline
 \multicolumn{2}{|l|}{thread}                           & 37  \\ \hline
 \multicolumn{2}{|l|}{scheduler}                        & 40  \\ \hline
 \multicolumn{2}{|l|}{rr}                               & 41  \\ \hline
 \multicolumn{2}{|l|}{delta\_list}                      & 56  \\ \hline
 \multicolumn{2}{|l|}{mutex}                            & 8   \\ \hline
 \multicolumn{2}{|l|}{awkernel\_lib}                     & 8 \\ \hline
 \multicolumn{2}{|l|}{verification}                     & 38 \\ \hline
 \multicolumn{2}{|l|}{rust\_primitives}                 & 111  \\ \hline \hline
 \multicolumn{2}{|l|}{Total}                            & 1009 \\ \hline
 \end{tabular}
 \caption{Lines of Code in the Code Review Model\label{tbl:loc}}
 \end{table}
 \subsection{Model Checking Results}
 \label{ssec:model-checking-result}

We executed model checking on a machine equipped with an AMD EPYC 7H12 2.6GHz CPU and 4TB of DDR4/3200 SDRAM. We set the number of tasks to three, which is sufficient for checking preemption and resumption across different CPU cores.

When two of these three tasks were configured as Heavy Tasks, the model checking ran out of memory, even with 4TB available. 
Although we identified the bug intentionally embedded in the model before the out-of-memory error, we could not complete the entire run with this configuration.

Model checking could be completed when we configured the first generated task as a Heavy Task and the other two as Light Tasks, with the first task generated immediately and the others spawned at arbitrary times by existing tasks.
This configuration required 570.6GB of memory and 1.43 hours.

Instead of using Linear Temporal Logic (LTL) formulae to verify conditions such as whether each task eventually reaches a \code{Terminated} state and remains there without re-entering a \code{Running} state, we implemented separate monitoring processes to observe and confirm these state transitions.

\subsection{Runtime Verification Results}
We executed Awkernel on QEMU and encountered the error shown in Figure~\ref{fig:rv-result} within a few seconds. For instance, the error message in Figure~\ref{fig:rv-result} indicates that we missed a case where a \code{Running} task is in the scheduling queue and dequeued by another non-primary core. Although the implementation is designed to check that a task is not in the \code{Running} state before it starts and to ignore it if it is, this function call was missed in the code review model. Hence, it revealed a gap between the model's assumptions and the actual behavior of the implementation.

Through continuous execution of runtime verification, we observed the actual state transitions. We discovered a bug where a \code{Preempted} task occasionally resumes running without re-entering the \code{Running} state as it should.

\begin{figure}[tbp]
  \centering
  \begin{lstlisting}
  [1389 DEBUG] rv/src/model.rs:69: Running(false) -- PollPending --> Waiting(false)
  [1390 DEBUG] rv/src/model.rs:69: Runnable(false) -- GetNext --> Running(false)
  [1457 DEBUG] rv/src/model.rs:69: Running(false) -- SetNeedSched --> Running(true)
  [1368 DEBUG] rv/src/model.rs:69: Running(true) -- Preempt --> Preempted(true)
  [1368 DEBUG] rv/src/model.rs:69: Runnable(false) -- GetNext --> Running(false)
  [1369 ERROR] kernel/src/nostd.rs:10: panic: panicked at awkernel_async_lib/src/task.rs:350:22:
  called `Result::unwrap()` on an `Err` value: "Error: an invalid transition is found.\r\n  current state: Running(false)\r\n  symbol: GetNext"
  \end{lstlisting}
  \caption{The Output When Runtime Verification is Conducted\label{fig:rv-result}}
\end{figure}

 \subsection{Found Bugs}
\subsubsection{Preemption Timing}
\label{ssec:mc-found-bugs}

The round-robin scheduler periodically checks the elapsed time for each running task since it became \code{Running}. If the scheduler finds a task whose elapsed time has exceeded the time quantum, it will send an IPI to the core running that task. However, these two operations---checking the elapsed time and executing preemption---are not atomic, which caused a bug.

The expected behavior is as follows:
\begin{enumerate}
  \item The scheduler checks the elapsed time of task $\alpha$ running on core 2.
  \item The scheduler decides to perform preemption on core 2.
  \item The scheduler sends an IPI to core 2.

Instead of using Linear Temporal Logic (LTL) formats to verify conditions such as whether each task eventually reaches a \code{Terminated} state and whether such a task will never enter a \code{Running} state again, we created processes for monitoring states and confirmed them.
  \item Core 2 receives the IPI and starts preemption.
  \item Task $\alpha$ changes its state from \code{Running} to \code{Preempted} and is moved back to the scheduling queue.
\end{enumerate}

However, due to the lack of atomicity, the following sequence could occur before the fix:
\begin{enumerate}
  \item The scheduler checks the elapsed time of task $\alpha$ running on core 2.
  \item The scheduler decides to perform preemption on core 2.
  \item \textit{Task $\alpha$ finishes, and the next task $\beta$ starts running on core 2}.
  \item The scheduler sends an IPI to core 2.
  \item Core 2 receives the IPI and starts preemption.
  \item Task $\beta$ changes its state from \code{Running} to \code{Preempted} even though task $\beta$ has just started.
\end{enumerate}

To fix this bug, we use a flag, \code{need\_sched}, which is defined in the task information structure. This flag was introduced to prevent the issue shown in section~\ref{para:running-example}. Before sending an IPI, the scheduler sets the \code{need\_sched} flag for the target task. With this modification, we can avoid the bug mentioned above.

If the scenario above occurs, preemption is canceled as follows:
\begin{enumerate}
  \item The scheduler checks the elapsed time of task $\alpha$ running on core 2.
  \item The scheduler decides to perform preemption on core 2.
  \item \textit{The scheduler sets the \code{need\_sched} flag of task $\alpha$}.
  \item \textit{Task $\alpha$ finishes, and the next task $\beta$ starts running on core 2}.
  \item The scheduler sends an IPI to core 2.
  \item Core 2 receives the IPI and starts preemption.
  \item \textit{Since the \code{need\_sched} flag of task $\beta$ is not set, preemption is canceled, and task $\beta$ resumes}.
\end{enumerate}

This seems to resolve the bug. However, the flag had already been used for \code{Waiting} tasks to address the issue shown in section~\ref{para:running-example}. Therefore, we had to ensure that no new issues would arise from this modification. It is difficult to be certain through human review alone due to concurrency. Hence, we used model checking for the code review. By employing model checking, we gained confidence and applied the above fix.
 
\subsubsection{Task Transition}
\label{ssec:rv-found-bugs}
We identified anomalous behavior in Awkernel through runtime verification. When tasks run, they should be in the \code{Running} state. However, during runtime verification, we observed cases where tasks sometimes resumed execution while still marked as \code{Preempted}. With the assistance of model checking, we traced the path leading to this error and implemented a fix.

Additionally, another issue was discovered during this process. While tracing the task transition states with runtime verification, we found that the \code{state} variable did not entirely reflect the task state. Each task had a flag called \code{in\_queue}, and the value of this flag affected the transition destination. Even when the \code{state} variable had the same value, the transition destination changed depending on the value of the \code{in\_queue} flag. Based on this observation, we revised the state transitions to align more closely with typical operating system behavior.

\section{Lessons Learned}
\label{sec:lessons-learned}
\subsection{Effort}
Implementing Awkernel and applying formal methods to it has been performed by others. The person who applied formal methods to Awkernel could ask the developer at anytime. Additionally, although the person who applied model checking knew the foundations of Promela, he needed to gain experience writing Promela for medium/large-scale software such as Awkernel. Under these conditions, the runtime verification took two weeks, and the model checking took three months.

In the runtime verification part, we first created an automaton that shows the transitions of the task states. Then, we embedded it into Awkernel as the monitor. This was quickly completed after consulting with the developer to understand the sequence of operations in the code. Subsequently, we executed the runtime verification, and when we found transitions not present in the model, we compared the model with the reports from the monitor, revised the model, and re-executed the process several times. After repeating this process and identifying the problematic parts of the implementation, we reported them to the developer. This entire sequence took two weeks.

In the model checking part, due to the semantic gap between Rust and Promela, it took three months, which is longer than the runtime verification. For instance, one semantic gap is that Rust can have a function called automatically when a variable's ownership is dropped, similar to a destructor in object-oriented languages. However, Promela lacks such a feature. Hence, we had to check every scope of variables and manually describe each drop function in Promela.

Moreover, to mimic preemption behavior, we assigned one Promela process to one thread context, but the description of this part became complex. For example, there was a bug where multiple processes representing the same thread context were created. However, a part we approximated contains numerous interleavings, making it difficult to identify the cause of such bugs.

\subsection{Semantic Gap}
Awkernel is implemented in Rust. Due to concepts like ownership in Rust, there is a semantic gap that cannot be overlooked with Promela. Therefore, theoretically speaking, even if we manually convert Rust code to Promela very carefully, we cannot say that all semantics are accurately translated, nor can we guarantee that all behaviors occurring in the implementation are covered by the Promela description.

However, in terms of using model checking for code review while running the code rather than for verification, this method of manually converting Rust code to Promela was valuable. In practice, using this method allowed us to verify whether the behavior was as intended more quickly and confidently than performing a completely manual code review.

\subsection{Assumptions}
When using model checking for the review, we should recognize that we can easily overlook implicit assumptions behind the system. Hence, even though model checking exhaustively verifies described behaviors, it must be noted that behaviors not described are not checked. 

Runtime verification is helpful in minimizing such oversights. In fact, we discovered two overlooked behaviors by performing runtime verification. These behaviors involved state transitions of tasks due to function calls related to drivers, which were outside the target of the Promela description. The oversight occurred because we assumed no function calls would affect task scheduling from outside the target files.

It is impossible to account for all events without omissions in the development of complex software. Therefore, it is crucial to remain aware of the potential for such oversights and to use various methods to minimize these omissions as much as possible.

\subsection{Unexpected Volume of Interleavings}
The code review model written in Promela consists of 1009 lines, as shown in Section~\ref{ssec:loc}.
While it is not a small model, it is also not excessively large. However, performing model checking on it requires more than 4TB of memory, especially when considering preemption. We have made significant efforts to reduce the number of interleavings in the model but have not yet found an effective solution. Despite the simplicity of the scheduling mechanism, the number of possible execution paths is enormous, contributing to a high number of interleavings.

When writing concurrent programs, we may not actively consider it, but we are dealing with a tremendous number of
potential execution paths. This can make missing specific paths that may lead to bugs easy. In this case, model checking will be helpful.

Although we have not fully resolved the state explosion problem, we would like to share a few strategies we have used that we find effective. First, grouping assignments and \code{printf} statements within \code{atomic} blocks is important.
Typically, the specific ordering of these statements does not need to be considered, and if left unordered, they can contribute to the state explosion problem.
Second, under-approximation is useful. Indeed, we under-approximated the tasks by setting limits on the number of preemptions by the round-robin scheduler and the number of times each task enters the \code{Waiting} state.  This approach allows us to divide the search space for model checking, reducing memory usage and runtime.
Therefore, determining how to effectively under-approximate the target is crucial.

 \section{Conclusions}
 \label{sec:conclusion}
Developing operating systems of autonomous driving vehicles is challenging due to their concurrency and asynchronous nature. For such software, simple testing and code reviewing by humans are not enough; they can easily miss an interleaving that leads to bugs.

To mitigate the difficulty of tracing behaviors, we proposed a development method that introduces model checking to the process, called model checking-assisted code review.
Even though this method helps trace behaviors and identify non-trivial interleavings, faithfulness is limited due to the semantic gap between a programming language and the language used by a model checker. To complement the drawbacks of model checking-assisted code review, we also introduced runtime verification to the development process.

We applied the proposed method to Awkernel, an operating system for autonomous driving vehicles, and found two bugs that were difficult for humans to identify. The first bug was related to the timing of preemption, and the second was related to task transitions. We fixed these bugs and confirmed the correctness of the fixes through model checking and runtime verification.

\section{Acknowledgements}
This work was supported by JST, CREST Grant Number JPMJCR23M1, Japan.
This research is also based on results obtained from a project, JPNP21027, subsidized by the New Energy and Industrial Technology Development Organization (NEDO)
Green Innovation Fund Projects / Development of In-vehicle Computing and Simulation Technology for Energy Saving in Electric Vehicles.

\bibliographystyle{plain}
\bibliography{references}

\end{document}